\documentclass[12pt]{iopart}

\usepackage{iopams}

\usepackage{graphicx} 
\usepackage{multirow}

\begin{document}

\title{Curie-Weiss magnet -- a simple model of phase transition}
\author{M Kochma{\'n}ski$^1$, T Paszkiewicz$^2$, S Wolski$^2$}
\address{$^1$ University of Rzesz{\'o}w, Rejtana 16A, 35-310 Rzesz{\'o}w, Poland}
\address{$^2$ Rzesz{\'o}w University of Technology, ul. Powsta{\'n}c{\'o}w Warszawy 6, 35-959 Rzesz{\'o}w, Poland}
\ead{tapasz@prz.edu.pl} 

\begin{abstract}
The Curie-Weiss model is an exactly soluble model of ferromagnetism that allows one to study in detail the thermodynamic functions, in particular their properties in the neighbourhood of the critical temperature. In this model every magnetic moment interacts with every other magnetic moment. Because of its simplicity and because of the correctness of at least of some of its predictions, the Curie-Weiss model occupies an important place in the statistical mechanics literature and its application to information theory. It is frequently presented as an introduction to the Ising model or to spin glass models, and usually only general features of the Curie-Weiss model are presented. We discuss here properties of this model in a rather detailed way. We present the exact, approximate and numerical results for this particular model. The exact expression for the limiting magnetic field is derived. 
\end{abstract}

\vspace{2pc}
\submitto{\EJP}
\maketitle

\section{Introduction}
\label{sect:1}
A phase transition is the transformation of a thermodynamic system from one phase or state to another. During a phase transition of a given medium certain properties change, often discontinuously, as a result of some external conditions, such as temperature, pressure or the magnetic field. For example, a liquid may become gas upon heating to the boiling point, resulting in an abrupt change in volume. 

In the case of a ferromagnet one should predict the dependence of such quantities as the free and internal energy, entropy, heat capacity and magnetic susceptibility on temperature $T$ and magnetic induction $B$. Generally, this is very complicated task. Ma \cite{ma1976} stressed the distinction between the direct approach to the problem of phase transitions and the approach exploiting symmetries of the problem. Here we shall illustrate the former approach. This means calculations of physical properties of interest in terms of parameters given in a particular model, i.e. solving a model. The calculations may be done analytically or numerically; exactly or approximately. 

One of the simplest classical systems exhibiting phase transition has been introduced by Pierre Curie and then by Pierre Weiss in their development of simplified theory of ferromagnetism~\cite{kac1969,nishimori2001}. Recently it is called the mean field theory. The Curie-Weiss considered a set of magnetic moments interacting with their nearest neighbours. They replaced the actual interactions experienced by each magnetic moment with the mean interaction given by the mean magnetization. With growing number of nearest neighbours the mean field theory becomes a better approximation. One can enlarge the number of nearest neighbours considering magnetic moments in spaces of higher dimensions (cf. ref.\cite{binney1992}). Mark Kac considered a model where every magnetic moment interacts with every other magnetic moment, and called it Curie-Weiss (CW) model \cite{kac1969}. Gould and Tobochnik called this model the fully connected Ising model \cite{gould2010}.

Although the Curie-Weiss model leads to the same results as the mean field theory for the behaviour of the system in vicinity of critical temperature, generally the thermodynamic functions calculated in the frame of these models are different. 
  
We shall focus on the Curie-Weiss model of a magnet. The Curie-Weiss model is an exactly soluble model of ferromagnetism that allows one to study in detail the behaviour of thermodynamic functions. Since not all predictions of this model agree with experiments, other models must be considered. However, because of its simplicity and because of the correctness of at least some of its predictions, the classical Curie-Weiss model occupies a central place in the statistical mechanics literature. It is frequently presented as an introduction to the Ising model or to spin glass models \cite{nishimori2001,adams2006,reichl1980}. However, in these references only general features of the Curie-Weiss model are discussed. This is why we shall discuss here properties of this model in a rather detailed way. We present the exact, approximate and numerical results for this particular model. 

We hope that this paper will be of use to such readers as graduate and postgraduate students as well as beginning research workers. We expect also that teachers might find our paper interesting enough to incorporate it in their course thereby introduce the students to the example richer than the mean field theory. 
\section{The Curie-Weiss model}
\label{sect:2}
Let us call the set of integers from $1$ to $N$ a lattice, and its element $i$ a site. We assign a variable $s_i$ (the Ising spin) to each site. The Ising spin is characterized by the binary value: $+ 1$ if microscopic magnetic moment is pointing up or $-1$ if it is pointed down. Particles with the Ising spins interact via Hamiltonian 
\begin{equation}
H_{int}=-\frac{J}{N}\sum\limits_{1\le i<j\le N}{{s}_{i}{s}_{j}} .
\label{eq:18.1}
\end{equation}
The constant $J$ is positive. The interaction energy of all pairs of spins of the Curie-Weiss magnet is the same and their interaction depends on $N$. The normalization by $1/N$ makes $H_{int}$ a quantity of the order N, i.e. an extensive quantity. The underlying assumption of an infinite-range interaction is clearly unphysical. The Hamiltonian \eref{eq:18.1} does not depend on dimension of the space which Curie-Weiss magnet is occupying.

The magnetic moment of a particle is proportional to the spin $\mu_{i}=\mu s_{i}$, where $\mu$ is the magnetic moment. In an applied magnetic field with the magnetic induction vector $\mathbf B$ particles with magnetic moments being parallel or antiparallel to $\mathbf B$, acquire the energy 
\begin{equation}
H_{f}=-\mu B\sum\limits_{i=1}^{N}s_{i}\,.
\label{eq:18.2}
\end{equation}  
The complete Hamiltonian consists of two terms 
\begin{equation}
H=-\frac{J}{N}\sum\limits_{1\le i<j\le N}{s}_{i}{s}_{j}-\mu B\sum\limits_{i=1}^{N}s_{i}\,.
\label{eq:18.3}
\end{equation}
The Hamiltonian \eref{eq:18.3} does not change if we reverse signs of all spins ${{s}_{i}}\to -{{s}_{i}}\,\,(i=1,2,\ldots ,N)$ and the direction of the induction vector $\mathbf{B}\to -\mathbf{B}$ 
\begin{equation}
H\left( {{s}_{1}},\ldots ,{{s}_{n}};\mathbf{B} \right)=H\left( -{{s}_{1}},\ldots ,-{{s}_{n}};-\mathbf{B} \right). 
\label{eq:symm-H}
\end{equation}

Denote a particular configuration $\left(s_{1},s_{2},\ldots,s_{N}\right)$ by $\left\{s\right\}$. To each configuration $\left\{s\right\}$ there corresponds an energy $E\left( \left\{ s \right\} \right)=H\left( \left\{ s \right\} \right)$. 

The nature of phase transitions of magnetic systems is well understood. At temperature $0 \rm K$ magnetic systems, in particular the CW magnet, are in a lowest energy state with all spins being parallel. Thus, their magnetization $M$ is finite and our magnet is ferromagnetic. As temperature is increased from zero the thermal noise randomizes spins. A fraction of them become antiparallel. This disorder grows with raising temperature and a diminishing fraction of them points at the initial direction. At temperature $T_c$ -- the critical temperature, and beyond, magnetization vanishes and the material becomes paramagnetic. For $T$ above $T_c$, there must be macroscopically large regions in which a net fraction of spins are aligned up. However, their magnetization mutually compensates -- they cannot make a finite fraction of all regions agree. For $T$ just below $T_c$ the compensation is not complete and the small, but finite, fraction points in the same direction. 

For left-hand vicinity of $T_c$ thermodynamic functions depend on the dimensionless parameter $t=\left(T-T_{c}\right)/T_{c}$ and consists of terms regular in $t$ and singular in it. The singular terms depend on powers of $\left|t\right|^{-1}$. These powers are called the critical indexes (or critical exponents) and are defined for $B=0$ and $t\rightarrow 0$. The specific heat $c$ per one particle and the magnetization and the magnetic susceptibility $\chi_T$ for the paramagnetic phase are characterized respectively by critical indexes $\alpha$, $\beta$ and $\gamma$ 
\begin{equation}
c \sim {\left| t \right|^{ - \alpha }}\,,\,\,{m \sim {\left| t \right|^{ - \beta }}}\,,\,\, {\chi _T} \sim {\left| t \right|^{ - \gamma }}\,.
\label{eq:4}
\end{equation} 
As we shall show these critical indexes characterizing the critical behaviour of both phases are the same. We shall calculate these sets of critical indexes, as well as the dependence of the internal energy and entropy on $\left|t\right|$, for ferromagnetic and paramagnetic phases of the Curie-Weiss magnet. 

In the presence of the magnetic field the dependence of magnetization on the magnetic field at $\left|t\right|=0$ is characterized by the critical index $\delta$
\begin{equation}
m \sim B^{1/\delta}\,.
\label{eq:5}
\end{equation}
In the following, in place of temperature we will use $\theta=k_{B}T$, $k_B$ being the Boltzmann constant.

\section{Calculation of free energy}
\label{sc:3}
Since $s_{i}^{2}=1$ the Hamiltonian \eref{eq:18.1} can be written as 
 \begin{equation}
{H}_{int}=-\frac{J}{2N}{{\left( \sum\limits_{i=1}^{N}{{{s}_{i}}} \right)}^{2}}+\frac{J}{2}.
\label{eq:18.5}
\end{equation}

The partition function is defined as usual as \cite{huang1987} 
\begin{equation}
{Z_N} = \sum\limits_{\left\{ s \right\}} {{e^{ - E\left( {\left\{ s \right\}} \right)/\theta }}} \,.
\label{eq:18.4}
\end{equation}
The summation is performed over all $2^N$ configurations ${\left\{ s \right\}}$. Note that a different method calculating spin configurations can be used \cite{gould2010}.

Introduce two dimensionless quantities $K=J/\theta$ and $h=\mu B/\theta$. Then, the partition function \eref{eq:18.4} can be written as
\begin{eqnarray}
\fl
  {Z_N}\left( \theta  \right) = \sum\limits_{\left\{ s \right\}} {\exp \left[ {\frac{K}{{2N}}{{\left( {\sum\limits_{i = 1}^N {{s_i}} } \right)}^2} - \frac{K}{2} + h\sum\limits_{i = 1}^N {{s_i}} } \right]}  =   \nonumber \\
   = {e^{ - \frac{K}{2}}}\sum\limits_{\left\{ s \right\}} {\exp \left[ {{{\left( {\sqrt {\frac{K}{{2N}}} \sum\limits_{i = 1}^N {{s_i}} } \right)}^2} + h\sum\limits_{i = 1}^N {{s_i}} } \right]} \,.
\label{eq:18.6}
\end{eqnarray}

Thermodynamic functions can be obtained via the Helmholtz free energy $F_{N}\left(\theta,B\right)$ \cite{huang1987}
\begin{equation}
F_{N}\left(\theta,B\right)=-\theta\ln Z_{N}\left(\theta,B\right).
\label{eq:6}
\end{equation}
The derivative of $F_{N}\left(\theta,B\right)$ \eref{eq:6} with respect to $B$ gives
\begin{equation}
\frac{\partial F_{N}\left( \theta ,B \right)}{\partial B}=-\mu\left\langle \sum\limits_{i=1}^{N}{{{s}_{i}}} \right\rangle =-\mu \sum\limits_{i=1}^{N}{\left\langle {{s}_{i}} \right\rangle } ,
\label{eq:1-st derv}
\end{equation}
where $\left\langle s_i\right\rangle$ means the mean value of $s_i$ calculated with the canonical distribution function \cite{huang1987}. 

The minus derivative of free energy \eref{eq:1-st derv} defines magnetization $M$ of a magnet \cite{huang1987}, thus 
\begin{equation}
M=\mu \sum\limits_{i=1}^{N}{\left\langle {{s}_{i}} \right\rangle }.
\label{eq:magnetization}
\end{equation}

The second derivative of $F_{N}\left( \theta ,B \right)$ with respect to $B$ is positive \cite{reif1967} 
\begin{equation}
\frac{{{\partial }^{2}}F_{N}\left( \theta ,B \right)}{\partial {{B}^{2}}}=\frac{\mu^{2}}{\theta}\left\{\left\langle \left( \sum\limits_{i=1}^{N}{{{s}_{i}}} \right)\left( \sum\limits_{j=1}^{N}{{{s}_{j}}} \right) \right\rangle -{{\left\langle \sum\limits_{i=1}^{N}{{{s}_{i}}} \right\rangle}^{2}}\right\}>0 .
\label{eq:2-nd-deriv}
\end{equation}
Since $\left[{{\partial }^{2}}F\left( \theta ,B \right)/{\partial {{B}^{2}}}\right]_\theta$ defines the magnetic susceptibility $\chi_T$ (cf. Eq.~\eref{eq:18.20}), we conclude that the magnetic susceptibility is positive. Therefore, the change of induction $\delta B$ increases the magnet energy by $\delta B\cdot \chi_{T} \cdot\delta B>0$. The CW magnet fulfills the condition of thermodynamic (macroscopic) condition of stability \cite{reichl1980}. 

A simple evaluation of the partition function \eref{eq:18.6} is precluded only by square of magnetization in the exponential. One can get rid of this square using the Gaussian linearization of the form 
\begin{equation}
{e^{{a^2}}} = \frac{1}{{\sqrt {2\pi } }}\int\limits_{ - \infty }^{ + \infty } {d\xi {e^{ - \xi }}^{^2/2 + \sqrt 2 a\xi }} \, .
\label{eq:18.7}
\end{equation}
In the present case $a=\sqrt{K/\left( 2N \right)}\sum\limits_{i=1}^{N}{{{s}_{i}}}$. Now, the partition function factors with respect to individual summations over the state $s_i$
\begin{eqnarray}
\fl
{Z_N} = \frac{{{e^{ - \frac{K}{2}}}}}{{\sqrt {2\pi } }}\sum\limits_{{s_1} =  - 1}^{ + 1} {\sum\limits_{{s_2} =  - 1}^{ + 1} { \ldots \sum\limits_{{s_N} =  - 1}^{ + 1} {} } }  \nonumber \\
  \int\limits_{ - \infty }^{ + \infty } {d\xi {e^{ - {\xi ^2}/2}}{e^{\sqrt {K/N} \,{s_1}\,\xi  + h{s_1}}}{e^{\sqrt {K/N} \,{s_2}\xi  + h{s_2}}}}  \ldots {e^{\sqrt {K/N} \,{s_N}\xi  + h{s_N}}} =  \nonumber \\
   = \frac{{{2^N}{e^{ - K}}}}{{\sqrt {2\pi } }}\int\limits_{ - \infty }^{ + \infty } {d\xi } {e^{ - {\xi ^2}/2}}{\left[ {\cosh \left( {\sqrt {K/N} \,\xi  + h} \right)} \right]^N}\,. 
\label{eq:18.8}
\end{eqnarray}
Performing the change of variable $\sqrt{K/N}\xi =Ky$ we get 
\begin{equation}
{Z_N} = 2^{N}{\left( {\frac{{KN}}{{2\pi }}} \right)^{1/2}}{e^{ - \frac{K}{2}}}\int\limits_{ - \infty }^{ + \infty } {dy} \Phi_{K,y}\left(y\right)\,,
\label{eq:18.9}
\end{equation}
where 
\begin{equation}
{\Phi_{K,h}}\left( y \right) = {e^{ - K{y^2}/2}}\cosh \left( {Ky + h} \right)\,.
\label{eq:18.10}
\end{equation}
In addition to $y$ the function $\Phi_{K,h}\left( y \right)$ depends on two dimensionless parameters $K$ and $h$, i.e. on $\theta$ and $B$. 

Free energy per one particle is proportional to $\left( \ln {{Z}_{N}} \right)/N$. Since we are interested in analyzing the system in the large size limit 
\[
\mathop {\lim }\limits_{N \to \infty } \frac{{\ln {Z_N}}}{N} = \mathop {\lim }\limits_{N \to \infty } \ln \left( {Z_N^{1/N}} \right) = \ln \left( {\mathop {\lim }\limits_{N \to \infty } Z_N^{1/N}} \right)\,.
\]
Using $Z_{N}$ \eref{eq:18.9} we obtain free energy per particle $f\left(\theta,B\right)$
\begin{eqnarray}
\fl
   - \frac{{f\left( {\theta ,B} \right)}}{\theta } = \mathop {\lim }\limits_{N \to \infty } \frac{1}{N}\ln \left( {{e^{ - \frac{K}{2}}}\sqrt {KN/2\pi } } \right) + \ln2 + \nonumber \\
   + \ln \mathop {\lim }\limits_{N \to \infty } {\left\{ {\int\limits_{ - \infty }^{ + \infty } {dy} \left[ {{\Phi _{K,h}}\left( y \right)} \right]{\,^N}} \right\}^{1/N}}\,. 
\label{eq:18.11}
\end{eqnarray}

In order to obtain the explicit form of the function $f\left(\theta, B\right)$ we use the Laplace theorem \cite{polya1976}. 

Theorem. Let functions $\varphi \left( y \right)$ and $\psi \left( y \right)$ will be continuous and positive in a range $c\le y\le d$, then 
\begin{equation}
\mathop {\lim }\limits_{n \to \infty } {\left\{ {\int\limits_c^d dy {\varphi \left( y \right){{\left[ {\psi \left( y \right)} \right]}^n}} } \right\}^{1/n}} = \mathop {\max }\limits_{c \le y \le d} \psi \left( y \right)\,.
\label{eq:Laplace}
\end{equation}

For $\varphi \left( y \right)=1$ and $\psi \left( y \right)=\Phi _{K,h}\left(y\right)$ in the limit $N\rightarrow \infty$ this theorem yields 
\begin{equation}
- \frac{{f\left( {\theta ,B} \right)}}{\theta } = \ln \mathop {\max }\limits_{ - \infty  \leq y \leq \infty } {\Phi _{K,h}}\left( y \right)+\ln 2\,. 
\label{eq:18.12}
\end{equation}

Let us introduce a function of $y$ related to free energy 
\begin{equation}
{{f}_{\theta ,h}}\left( y \right)=-\theta \left[ \ln 2+\ln {{\Phi }_{\theta ,h}}\left( y \right) \right].
\label{eq:f_Kh}
\end{equation}
According to Eq. \eref{eq:18.12} to find the dependence of free energy on thermodynamic variables $\theta$ and $B$ one should find extreme points of $f_{K,h}\left(y\right)$. For these points $\left(df_{K,y}\left(y\right)/dy\right)_{\theta,B}=0$. Calculating the derivative of $\Phi_{K,h}\left( y\right)$ \eref{eq:18.10} we find 
\begin{eqnarray*}
{{\left( \frac{\partial f}{\partial y} \right)}_{\theta ,B}}=-\frac{{{\theta }K}}{{{\Phi }_{K,h}}\left( y \right)}\left[ \tanh \left( Ky+h \right)-y \right]{{\Phi }_{K,h}}\left( y \right)=
\\=-{{\theta }}\left[ \tanh \left( Ky+h \right)-y \right].
\end{eqnarray*}
Thus, the variable $y$ obeys the equation 
\begin{equation}
y=\tanh \left( Ky+h\right).
\label{eq:18.13}
\end{equation}

For various values of $\theta$ ($K$) and $B$ ($h$) solutions of this equation provide the function $y=y\left(\theta,B\right)$ of state variables. Therefore, the function $\Phi \left( \theta ,B \right)=\mathop {\max} \limits_{-\infty \le y\le \infty } \,{{\Phi }_{K,h}}\left( y \right)$  is a composite function of $\theta$ and $B$, namely $\Phi_{K,y}\left(y\left(\theta, B\right)\right)$ also depends implicitly on these two state variables. Now Eq. \eref{eq:18.12} can be rewritten in the form
\[
f\left( \theta, B \right)=-\theta\ln 2 -\theta\ln \Phi\left(\theta, B\right),
\]
or 
\begin{equation}
f\left( \theta ,B \right)=-\theta \ln 2-\theta \ln \left\{ {{e}^{-K{{y^{2}}\left(\theta, B\right)}/2}}\cosh \left[ Ky\left( \theta ,B \right)+h \right] \right\} . 
\label{eq:18.15}
\end{equation}

Using two familiar identities \cite{bronshtein2004} $d\tanh x/dx=\cosh^{-2}x$ and 
\begin{equation}
\cosh^{2}x=\left(1-\tanh^{2} x\right)^{-1}=\left(1-y^{2}\right)^{-1}, 
\label{eq:identity-1}
\end{equation}
we calculate the second partial derivative of $f$ with respect to $y$ 
\begin{equation}
{{\left. {{\left( \frac{{{\partial }^{2}}f}{\partial {{y}^{2}}} \right)}_{\theta ,h}} \right|}_{y=y\left( \theta ,B \right)}}=-{\theta }K\left\{ K\left[ 1-{{y}^{2}}\left( \theta ,B \right) \right]-1 \right\}.
\label{eq:2-nd-deriv-y}
\end{equation}
In the Appendix we show that there exist solutions of Eq. \eref{eq:18.13} for which this derivative is positive, hence for them free energy is minimal. 

For the reversed magnetic field the solution of Eq. \eref{eq:18.13} is $-y$ 
\begin{equation}
-y=\tanh \left[ K\left( -y-h \right) \right].
\label{eq:anti-symm-y}
\end{equation} 
Therefore, free energy $f\left(\theta,B\right)$ \eref{eq:18.12} is an even function of $B$ (and $h$)
\begin{eqnarray}
	\fl
	-\frac{f\left(\theta,-B\right)}{\theta}=\ln 2 + \ln\left\{e^{ - K{\left(-y\left(\theta,B\right)\right)^2}/2}\cosh \left[ {-K\left( -y\left(\theta,B\right)\right) - h} \right]\right\}=\nonumber \\
	=-\frac{f\left(\theta,B\right)}{\theta}\,.
\end{eqnarray}  
These properties of free energy and solutions of Eq. \eref{eq:18.13} are the reason why in the following we will always have in mind the positive value of $B$. 

Calculating derivatives $\left(\partial y/\partial h\right)_{\theta}$ and $\left(\partial^{2}f/\partial y^{2}\right)_{K,h}$ one may show that the product of ${{\left( \partial y/\partial h \right)}_{\theta }}$ and ${{\left( {{\partial }^{2}}{{f}_{K,h}}\left( y \right)/\partial {{y}^{2}} \right)}_{K ,h}}$ is positive because 
\begin{equation}
{{\left( \frac{\partial y}{\partial h} \right)}_{\theta }}{{\left( \frac{{{\partial }^{2}}{{f}_{K,h}}\left( y \right)}{\partial {{y}^{2}}} \right)}_{K ,h}}=\frac{\theta K}{{{\cosh }^{2}}\left( Ky+h \right)}. 
\label{eq:identity}
\end{equation} 
 
Introduce here $K_{c}=1$ -- the critical value of the parameter $K$. From the definition of $K$ it is seen that the critical value of $\theta$ is $\theta_{c}=J$. 
\section{Free energy of CW magnet in the absence of the magnetic field}
\label{sc:4}
\begin{figure}[ht]
	\centering
		\includegraphics[width=0.90\textwidth, angle=0]{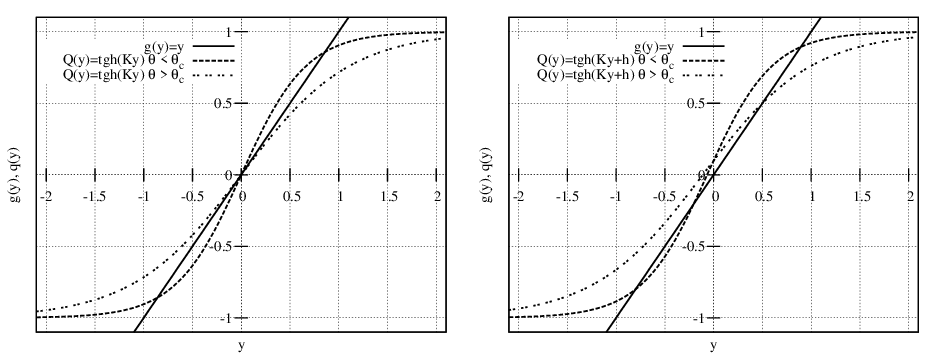}
	\caption{Graphical solution of Eq. \eref{eq:18.13}. The full line represents the function $g(y)=y$. The dashed lines: $\theta<\theta_{c}$ ($K$=1.5), the dotted lines: $\theta\geq \theta_c$ ($K$=0.9). Left panel: $h=0$ ($B=0$). Right panel: $h=0.1$.}
	\label{fig:18.1}
\end{figure}

Suppose that $B=0$ ($h=0$). If we plot $g\left(y\right)=y$ and $Q\left(y\right)=\tanh\left(Ky\right)$ as functions of $y$, the points of intersection determine solutions of Eq. \eref{eq:18.13}. Referring to Fig. \ref{fig:18.1} (left panel), we have to make distinction between cases. If $\theta > \theta_c$ ($\theta>J$) the slope of of the function $Q\left(y\right)$ at the origin $K=J/\theta=\theta_{c}/\theta <1$ is smaller than the slope of linear function $g\left(y\right)=y$, which is 1, thus these graphs intersect only at the origin. It is easy to check that for this solution the second derivative of free energy \eref{eq:18.12} is positive (cf. Appendix). Therefore, the extreme indeed is a minimum (cf. Fig. \ref{fig:18.2}). 

On the other hand when $\theta<\theta_c$ ($\theta<J$), the initial slope of $\tanh(Ky)$ is larger than that of linear function, but since values of $\tanh$ function cannot take values outside the interval $\left(-1,+1\right)$, the two functions have to intersect in two additional, symmetric nonzero points $\pm y\left(\theta\right)$ (Fig. \ref{fig:18.1}). In this case in the Appendix we show that the second derivative of free energy is negative at the origin $y=0$, which means that there is a maximum at $y=0$ (cf. Fig. \ref{fig:18.2}). This derivative is positive at $y=\pm y\left(\theta\right)$. Free energy $f\left(\theta\right)$ attains minimal value at $y=\pm y\left(\theta\right)$, hence these solutions correspond to the thermodynamically stable states.

The parameter $K$ and temperature $\theta$ can be expressed in terms of $t=\left(\theta/\theta_{c}-1\right)$, namely 
\begin{equation}
\eqalign
   K=\frac{{{\theta }_{c}}}{\theta }=\frac{1}{1+t}, \\ 
  \theta =\left( 1+t \right){{\theta }_{c\,}}.  
\label{eq:theta-K-vs-t}
\end{equation}

\section{Magnetization and magnetic susceptibility of Curie-Weiss magnet}
\label{sc:5}
Consider the CW magnet when the magnetic field $B$ is brought back. Magnetization per particle, $m$, is a partial derivative of free energy $f\left(\theta,B\right)$ after $B$
\begin{eqnarray}
\fl
  m\left( \theta ,h \right)=-\left[\frac{\partial f\left(\theta ,h \right)}{\partial B}\right]_{\theta}=\mu \left[\frac{\partial \ln \Phi \left( \theta ,h \right) }{\partial h}\right]_{\theta}= \nonumber \\ 
  =\mu \left.\frac{\partial \ln {{\Phi }_{K,h}}\left( y\right)}{\partial h}\right|_{y=y_{m}}+\mu\left.\frac{\partial \ln {\Phi}_{K,h}\left(y\right)}{\partial y}\right|_{y=y\left(\theta,B\right)}\frac{\partial y\left(\theta,B\right)}{\partial h}\, . 
\label{eq:mu}
\end{eqnarray}
Since $\Phi_{K,h}\left(y\right)$ attains an extremum at $y_{m}\left(\theta,B\right)$, the second term on right hand side of \eref{eq:mu} vanishes. The contribution of the first term yields the equation of state
\begin{equation}
m = \mu \tanh \left( {\frac{K}{\mu}m + h} \right)\,,
\label{eq:18.16}
\end{equation}
which has a closed analytic form. According to Eq. \eref{eq:18.13}, the study of $y$ is equivalent to the study of of magnetization $m$. 

\begin{figure}[ht]
	\centering
		\includegraphics[width=0.95\textwidth]{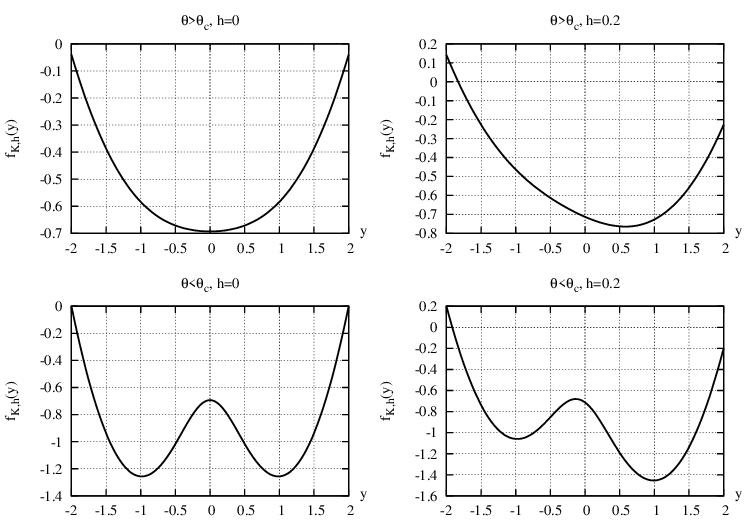}
	\caption{Dependence of function $f_{K,h}\left(y\right)$ \eref{eq:18.12} on $y$ for paramagnetic ($\theta\geq \theta_c$) and ferromagnetic ($\theta<\theta_c$) phases in the magnetic field ($h>0$) and when the magnetic field is absent ($h=0$).} 
	\label{fig:18.2}
\end{figure}

Consider the solutions of Eq. \eref{eq:18.13} for $B> 0$. When $\theta<\theta_c$ the plots of functions $g\left(y\right)=y$ and $Q\left(y\right)=\tanh\left(Ky+h\right)$ intersect in three non-symmetric and non-zero points (cf. right panel of Fig. \ref{fig:18.1}). For $h\ne 0$ Eq. \eref{eq:18.13} has two solutions if $\left| h \right|\le {{h}_{t}}$, where $h_{t}$ is some limiting value of $h$. This problem is discussed in Sects. \ref{sc:6} and \ref{sc:8}.

Only for positive value of $y=y\left(\theta,B\right)$ free energy $f$ attains the global minimum. To one of negative values there corresponds a local minimum, to the remaining a maximum (cf. Fig. \ref{fig:18.2}). The negative values of $y\left(\theta,B\right)$ (as well as $m\left(\theta, B\right)$) do not correspond to stable states and should be omitted. One should notice that negative values of $y$ for positive $h$ are not compatible with the symmetry \eref{eq:anti-symm-y} of the state equation. When $\theta\geq \theta_c$ the graphs of $g\left(y\right)$ and $Q\left(y\right)$ intersect at one point $y\left(\theta,B\right)>0$. For this value of $y\left(\theta,B\right)$ free energy has global minimum (cf. Fig. \ref{fig:18.2}). 

\begin{figure}[ht]
	\centering
		\includegraphics[width=0.90\textwidth]{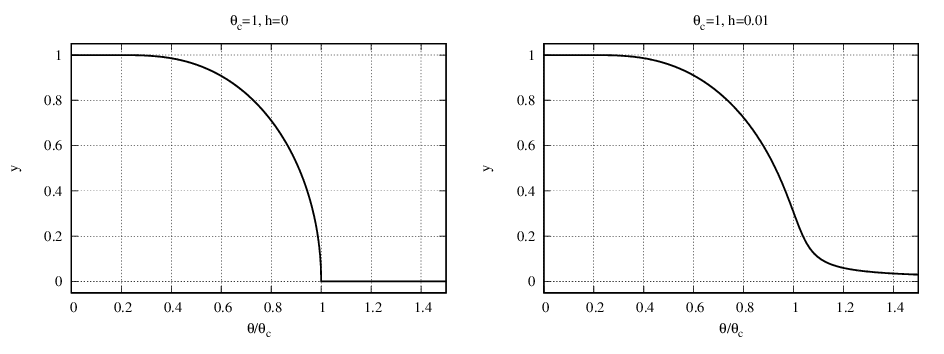}
	\caption{Dependence of of magnetic susceptibility on $\theta/\theta_c$. Left panel: $h=0$. Right panel: $h>0$.} 
	\label{fig:18.3}
\end{figure}

Consider Eq. \eref{eq:18.13} for small values of $\left|t\right|$ and $h=0$. We can expand $\tanh\left(Ky\right)$ into Taylor's series. Using Eq. \eref{eq:theta-K-vs-t} we obtain 
\begin{equation}
y\simeq \sqrt{3}\frac{{{\left( K-1 \right)}^{1/2}}}{{{K}^{3/2}}}\sim {{\left| t \right|}^{1/2}}. 
\label{eq:y-vs-abs-t}
\end{equation}
Hence, in agreement with Fig. \ref{fig:18.3}, we conclude that 
\begin{equation}
m \sim\left|t\right|^{1/2}.
\label{eq:m-vs-abs-t }
\end{equation}

When positive $B\rightarrow 0$ then $y\left(\theta,h\right)\rightarrow y\left(\theta\right)$ of Fig \ref{fig:18.3}. Thus, we note the existence of spontaneous magnetization $m\left(\theta\right)=\mu y\left(\theta\right)$
\begin{equation}
m\left( {\theta ,h = {0^ + }} \right) = \left\{ {\begin{array}{*{20}{c}}
{0\;\;\,\quad {\theta _c} \le \theta ,}\\
{{m_0}\quad {\theta _c} > \theta .}
\end{array}} \right.
\label{eq:27}
\end{equation}

Even after turning off the magnetic field, below critical temperature the system remains magnetized, depending on the sign of $B$ before its removal. The dependence of $y$ for $B=0$ on $\theta$ is singular. It is seen in Fig. \ref{fig:18.3} that at the point $\theta=\theta_c$ tangents of two branches of the curve are different. We note that the point $\theta=\theta_c$ is the boundary between the region of existence and nonexistence of magnetization, i.e. it is a critical point. We conclude that magnetization $m$ is the order parameter. 

In the presence of even an arbitrarily weak $\left(0\leq h \ll 1\right)$, the magnetic field magnetization $m$ does not vanish below and above critical point (cf. Fig. \ref{fig:18.3} the right panel). The external magnetic field lowers the symmetry of the paramagnetic phase. From the point of view of magnetization the difference between paramagnetic and ferromagnetic phases vanishes and the critical point ceases to exist. 

Consider the magnetic susceptibility
\begin{equation}
{{\chi }_{T}}\left( \theta ,B \right)=\left[\frac{\partial m\left( \theta ,B \right)}{\partial B}\right]_\theta\,.
\label{eq:18.20}
\end{equation} 
According to the definition \eref{eq:18.20} and Eq. \eref{eq:mu} the susceptibility is related to the second derivative of free energy with respect to induction
\begin{equation}
{{\chi }_{T}}\left( \theta ,B \right)=-\left[\frac{{{\partial }^{2}}f\left( \theta ,B \right)}{\partial {{B}^{2}}}\right]_{\theta}.
\label{eq:29}
\end{equation}
Differentiating both sides of Eq. \eref{eq:18.16} with respect to $B$ and solving the obtained equation, we obtain the general expression for the susceptibility 
\begin{equation}
{{\chi }_{T}}\left( \theta ,B \right)=\frac{\mu^2}{\theta }\frac{\left[1-y^{2}\left(\theta,B\right) \right]}{1-K\left[ 1-y^{2}\left(\theta,B\right)\right]}.
\label{eq:chi}
\end{equation}
Using Eq. \eref{eq:2-nd-deriv-y} the susceptibility can be written as
\begin{equation}
\chi_{T} \left( \theta ,B \right)=\frac{{{\mu }^{2}}}{\theta}\frac{1-{{y}^{2}}\left( \theta ,B \right)}{{{\left. {{\left( \frac{{{\partial }^{2}}{{f}_{K,h}}\left( y \right)}{\partial {{y}^{2}}} \right)}_{\theta ,h}} \right|}_{y=y\left( \theta ,B \right)}}}. 
\label{eq:chi-1}
\end{equation}
Since for a stable state the second derivative ${{\left. {{\left( {{\partial }^{2}}{{f}_{K,h}}\left( y \right)/\partial {{y}^{2}} \right)}_{\theta ,h}} \right|}_{y=y\left( \theta ,B \right)}}$ is nonnegative and $0\leq \left| y \right|\leq 1$, the susceptibility is nonnegative too. Notice that we succeed in linking together the macroscopic \eref{eq:2-nd-deriv} and microscopic \eref{eq:chi} stability conditions.

For the paramagnetic phase ($\theta\geq \theta_c$ ($t\geq 0$)) $y=0$, and in the vicinity of the critical temperature we obtain
\begin{equation}
\chi_{T} =\frac{{{\mu }^{2}}}{{{\theta }_{c}}\left( 1+t \right)}\frac{\left( 1+t \right)}{t}\simeq \frac{{{\mu }^{2}}}{{{\theta }_{c}}}{{t}^{-1}}.
\label{eq:chi-param}
\end{equation}
This means that the critical index $\gamma=1$.

The function ${\rm arctanh}y$ obeys the equation
\begin{equation}
{\rm arctanh}y=Ky+h\,.
\label{eq:31}
\end{equation}
We shall study solutions of Eq. \eref{eq:31} in vicinity of the critical temperature. For small $y$ one can expand ${\rm arctanh} y$ into Taylor's series \cite{bronshtein2004}. For $\theta<\theta_c$ ($t<0$) Eq. \eref{eq:31} reduces to
\begin{equation}
{{y}^{3}}-3\frac{\left|t\right|}{1-\left|t\right|}y-3h=0. 
\label{eq:18.21}
\end{equation}

In the ferromagnetic phase and in the absence of the magnetic field, the order parameter does not vanish
\begin{equation}
{{y}^{2}}=\frac{3\left| t \right|}{1-\left| t \right|}\,\,\left( t<0 \right)\,. 
\label{eq:18.26}
\end{equation}

When $B=0$ from Eq. \eref{eq:18.26} it follows that for ferromagnetic phase in vicinity of critical temperature $m\sim \sqrt{3\left|t\right|}$. Thus the critical index of magnetization is $\beta = 1/2$. 
Consider the susceptibility \eref{eq:chi} in ferromagnetic phase in the vicinity of the critical temperature. Using the expressions \eref{eq:theta-K-vs-t} and \eref{eq:18.26} for the ferromagnetic phase we obtain
\begin{equation}
{\lim_{B\to 0}}\,\chi_{T} \left( \theta ,B \right)\sim \frac{\mu^{2}}{\theta_c}\left|t\right|^{-1}\,,
\label{eq:18.25'}
\end{equation}
and according to the definition \eref{eq:4} for both phases the critical index of susceptibility is $\gamma =1$. 
\begin{figure}[ht]
	\centering
		\includegraphics[width=0.90\textwidth]{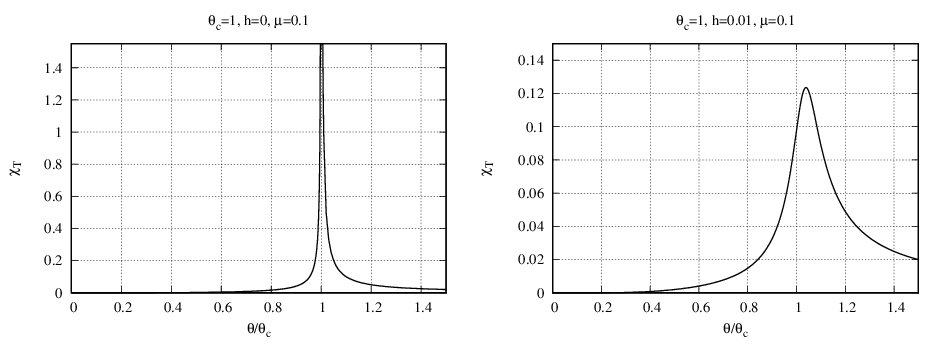}
	\caption{Dependence of the magnetization on $\theta/\theta_c$. Left panel: $h=0$. Right panel $h>0$.}
	\label{fig:18.4}
\end{figure}
We conclude that for both phases the magnetic susceptibility of Curie-Weiss magnet is divergent at the critical temperature. This singular behaviour is shown in Fig. \ref{fig:18.4}.

At the critical point $t=0$, hence from Eq. \eref{eq:18.21} it folows that $y=\sqrt[3]{3}\,{{h}^{1/3}}$. Therefore, according the definition \eref{eq:5}, the critical index for the critical isotherm $\delta=3$.

Continuous phase transitions occur when a new state of reduced symmetry develops continuously from the disordered (high temperature) phase. The ordered phase of CW magnet has lower symmetry than the symmetry \eref{eq:symm-H} of the Hamiltonian, thus the symmetry is spontaneously broken. There exist two equivalent symmetry related states of CW magnet with magnetization $+m$ i $-m$ respectively, with equal free energies. These states are macroscopically different, so thermal fluctuations will not bring them into contact in the thermodynamic limit. To describe the ordered state we introduced magnetization -- the macroscopic order parameter that describes the character and strength of the broken symmetry.
\section{Approximate theory -- the analysis of roots of the cubic equation for magnetization}
\label{sc:6}
Now we shall study the roots of the cubic equation \eref{eq:18.21} for the ferromagnetic phase. Since free energy is an even function of $B$ we assume that $h>0$. We shall study the so called the incomplete cubic equation
\begin{equation}
{{y}^{3}}+3\left(-p\right)y+2q=0\, ,
\label{eq:18.29}
\end{equation}
where for ferromagnetic phase $p=\left|t\right|/\left( 1-\left|t\right|\right),\,\,q=-3h/2$.

It is worthwhile to recall that assumptions proposed by Landau for an incompressible magnet result in free energy depending only on even powers of magnetization \cite{binney1992,reichl1980,als-nielsen1977} 
\[
f\left( m,T \right)={{f}_{0}}\left(T \right)+\alpha \left( T \right){{m}^{2}}+\frac{1}{2}\beta \left( T \right){{m}^{4}}. 
\]
This form of free energy also leads to a cubic equation for magnetization.

Introduce a characteristic value of the parameter $h$
\begin{equation}
{{h}_{t}}=\pm\frac{2}{3}\left| t \right|^{\frac{3}{2}},\,\, (\left|t\right|\ll 1).
\label{eq:h_t}
\end{equation}
The function $\Lambda_{app}\left(y\right)= y^{2}-3py$ has two extreme points at $y^{\left(extr\right)}_{\pm}\sim\pm \left| {t} \right|^{1/2}$. For these values of $y$ the parameter $h$ is equal to $\pm h_{t}$. Such method of finding the limiting value of magnetic field was used by Landau and Lifshits \cite{landau1980}.

Roots of Eq. \eref{eq:18.29} depend on the sign of the discriminant\cite{bronshtein2004} $D=\left(q^{2}-p^3 \right)$ 
\begin{equation}
D={{\left( \frac{3}{2}h \right)}^{2}}-{{\left( \frac{\left|t\right|}{1+\left|t\right|}\right)}^{3}}={{\left( \frac{3}{2} \right)}^{2}}\left( {{h}^{2}}-h_{t}^{2} \right) . 
\label{eq:18.30}
\end{equation}
If $D<0$ inequalities $-{{h}_{t}}<h<{{h}_{t}}$ hold. If $D>0$, then $h>h_{t}$ or $h<-h_{t}$. 

If $D<0$ all three roots are real
\begin{equation}
\eqalign{
y_{1}^{\left(<\right)}\left(t,h\right)=u_{<}\left(t,h\right)+v_{<}\left( t,h\right), \cr 
y_{2}^{\left( < \right)}\left( t,h\right)={{\varepsilon }_{1}}u_{<}\left( t,h\right)+{{\varepsilon }_{2}}v_{<}\left( t,h\right), \cr
y_{3}^{\left(<\right)}\left(t,h\right)={{\varepsilon }_{2}}u_{<}\left( t,h\right)+{{\varepsilon }_{1}}v_{<}\left( t,h\right), \cr
} 
\label{eq:<real-roots}
\end{equation}
where
\begin{equation}
\eqalign{
u_{<}\left( t,h\right)=\sqrt[3]{-q+i\sqrt{\left| D \right|}},\,\,\,v_{<}\left( t,h\right)=\left[{u}_{<}\left( t,h\right)\right]^{*}, \cr 
{{\varepsilon }_{1}}=\left( -1+i\sqrt{3} \right)/2,\,\,{{\varepsilon }_{2}}=\varepsilon _{1}^{*}\,. \\ 
}
\label{eq:symbols}
\end{equation}

Using relations \eref{eq:symbols} we can show that 
\begin{equation*}
\eqalign{
  {{y}^{\left(<\right)}_{1}}\left( t,h\right)=2{\rm Re}\;u_{<}\left(t,h\right), \cr 
  \left(y^{\left(<\right)}_{\sigma }\right)^{*}\left( t,h\right)={{y}^{\left(<\right)}_{\sigma }}\left( t,h\right)\,\,\left( \sigma =2,3 \right)\,.  
}
\end{equation*}

If $D>0$ (i.e. $h^{2}>{{h}^{2}_{t}}$) only one root is real 
\begin{equation}
{{y}_{>}}\left( t,h\right)={{u}_{>}}\left( t,h\right)+{{v}_{>}}\left(t,h\right), 
\label{eq:>root}
\end{equation}
where 
\begin{equation}
{{u}_{>}}\left(t,h \right)=\sqrt[3]{\frac{3h}{2}+\sqrt{D}},\,\,\,{{v}_{>}}\left(t,h \right)=\sqrt[3]{\frac{3h}{2}-\sqrt{D }}. 
\label{eq:> real-root}
\end{equation}
The two remaining two roots are complex and conjugated. 

If we combine Eqs. \eref{eq:h_t}-\eref{eq:> real-root} we obtain the functions $u_{<}\left(t,u\right)$, $v_{<}\left(t,u\right)$ and $u_{>}\left(t,u\right)$, $v_{>}\left(t,u\right)$ in the useful form 
\begin{equation}
\eqalign{
& {{u}_{>}}\left( t,h \right)={{\left( 3/2 \right)}^{1/3}}\sqrt[3]{h+\sqrt{{{h}^{2}}-h_{t}^{2}}},\,\,{{v}_{>}}\left( t,h \right)={{\left( 3/2 \right)}^{1/3}}\sqrt[3]{h-\sqrt{{{h}^{2}}-h_{t}^{2}}}, \cr
 & {{u}_{<}}\left( t,h \right)={{\left( 3/2 \right)}^{1/3}}\sqrt[3]{h+i\sqrt{h_{t}^{2}-{{h}^{2}}}},\,\,{{v}_{<}}\left( t,h \right)={{\left( 3/2 \right)}^{1/3}}\sqrt[3]{h-i\sqrt{h_{t}^{2}-{{h}^{2}}}}\,. 
} 
\label{new-parametriz}
\end{equation}
Consider functions $u_{>}\left(t,h\right)$ and $v_{>}\left(t,h\right)$ for negative $h$
\begin{equation*}
\eqalign{
{{u}_{>}}\left( t,-h \right)=\sqrt[3]{-1}\,\sqrt[3]{h-\sqrt{{{h}^{2}}-h_{t}^{2}}}=\sqrt[3]{-1}\,{{v}_{>}}\left( t,h \right),\,\, \cr 
{{v}_{>}}\left( t,-h \right)=\sqrt[3]{-1}\,\sqrt[3]{h+\sqrt{{{h}^{2}}-h_{t}^{2}}}=\sqrt[3]{-1}\,{{u}_{>}}\left( t,h \right).\,\,  
} 
\end{equation*}
Similar relations hold also for $u_<$ and $v_<$. Among three root of unity $\varepsilon_{1}$, $\varepsilon_{2}$ and $\varepsilon_{3}=-1$ only the latter root yields proper symmetry relation \eref{eq:anti-symm-y} 
\begin{equation}
y_{3}\left( t,-h \right)=-y_{3}\left( t,h \right).
\label{symm-rel}
\end{equation}

\begin{figure}[ht]
	\centering
		\includegraphics[width=0.50\textwidth]{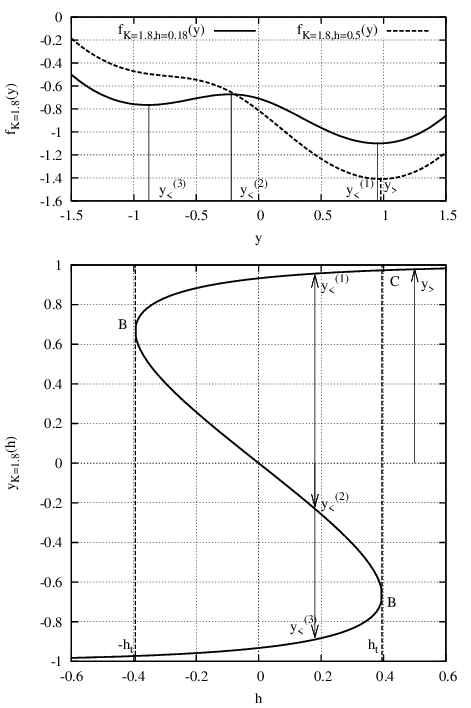}
	\caption{Plot of dependence of roots of Eq. (\ref{eq:18.29}) on $h$ for $h_{t}=0.4$. For $\left| h \right|>{{h}_{t}}$ the curve represents the function $y_{>}\left(t,h\right)$. In the interval $-{{h}_{t}}<h<{{h}_{t}}$ the plot consists of $y_{<}^{(j)}\left(t,h\right)$ ($j=1,2,3$)}
	\label{fig:18.5}
\end{figure}

In Fig. \ref{fig:18.5} we have plotted the dependence of roots \eref{eq:<real-roots} and \eref{eq:> real-root} on the magnetic field $h$. For positive values of $y_{<}^{(1)}$ and $-h_{t}<h<0$, as well as for negative values of $y_{<}^{(3)}$ and $0<h<h_t$, the signs of $h$ and these two roots do not agree. The symmetry \eref{eq:anti-symm-y} of equation of state is broken. On the line BB' (corresponding to the root $y_{<}^{(2)}$) the derivative ${{\left( \partial y\left( \theta ,h \right)/\partial h \right)}_{\theta }}<0$ is negative. From Eq. \eref{eq:identity} it follows that ${{\left( {{\partial }^{2}}{{f}_{\theta ,h}}\left( y \right)/\partial {{y}^{2}} \right)}_{\theta ,h}}$ is negative too. In this interval of values of $h$ free energy has maxima, therefore CW magnet is not in stable states. We conclude that for the root $y_{<}^{(2)}\left(t,h\right)$ free energy attains maximal values. 

For $0<h<h_t$ the root $y_{<}^{(1)}$ is positive. The remaining two roots are negative and $\left| y_{<}^{\left( 3 \right)} \right|< y_{<}^{\left( 1 \right)}$. Further $y_{<}^{\left( 2 \right)}\to y_{<}^{\left( 3 \right)}$ when $h\to h_{t}^{-}\equiv {{h}_{t}}-\varepsilon ;\,\,(0<\varepsilon <<1)$ and 
\[
{\lim_{h\to h_{t}^{-}} }\,{{\left( \frac{\partial y_{<}^{\left( 2 \right)}\left( \theta ,h \right)}{\partial h} \right)}_{\theta }}=-\infty.
\] 
From Eq. \eref{eq:identity} it follows that 
\[
{\lim_{h\to {{h_t}^{-}}} }\,{{\left( \frac{{{\partial }^{2}}{{f}_{\theta ,h}}\left( y \right)}{\partial {{y}^{2}}} \right)}_{\theta ,h}}=0.
\] 
This means that for $h=\pm h_t$ one should consider the third derivative of free energy with respect to $y$.

In the interval $-h_{t}<h<h_t$ the derivative of the root $y_{<}^{(3)}$ is positive, therefore the derivative ${{\left. {{\left( {{\partial }^{2}}{{f}_{\theta ,h}}\left( y \right)/\partial {{y}^{2}} \right)}_{\theta ,h}} \right|}_{y=y_{<}^{\left( 3 \right)}}}$ is also positive. This means that in this interval free energy has a (local) minimum. Free energy exhibits the global minimum for the root $y_{<}^{(1)}$ for $h_{t}>h>0$. This behaviour of free energy and of solutions of cubic equation is shown in Fig. \ref{fig:18.5}. 

For $h>h_t$ and $h<-h_t$ free energy has the global maximum for the real root $y_{>}\left(t,h\right)$ (cf. Fig. \ref{fig:18.5}). 

Note that as a result of approximation yielding Eq. \eref{eq:18.21} values of $\left|y\right|$ may exceed the limiting value 1. This means that we shall restrict ourselves to small values of $\left|t\right|$ and $h_t$. 

We shall point out that value $h_t$ \eref{eq:h_t} of parameter $h$ define such value $B_t$ of magnetic induction $B$, for which the value of induced magnetization $m_{ind}\sim \chi_{T} \cdot B_{t}$ ($B_{t}=\theta_{c}h_{t}/\mu$ with $h_t$ given by Eq. \eref{eq:chi-param}), is in accordance with Eq. \eref{eq:18.26}. 

If $h\ll h_t$ ($\left|t\right|\neq 0$) the magnetic field $B$ is week and does not influence the thermodynamic quantities characterizing the system. If $h\gg h_t$ the field $B$ is strong. If $t=0$ ($T=T_c$) all magnetic fields are strong. As we have shown, if $t\sim 0$ and field is strong, $m\sim h^{1/3}$. 

The parameter $h_t$ \eref{eq:h_t} divides the positive $h$-semiaxis into two parts. For $0<h\le h_t$ there exists three roots of Eq. \eref{eq:18.29} and to one of them there corresponds a global minimum of the free energy. For $h>h_t$ there exists one root corresponding to a minimum of free energy. 
\section{Properties of the internal energy, entropy and specific heat of the Curie-Weiss magnet}
\label{sc:7}
\begin{figure}[ht]
	\centering
		\includegraphics[width=0.80\textwidth, angle=0]{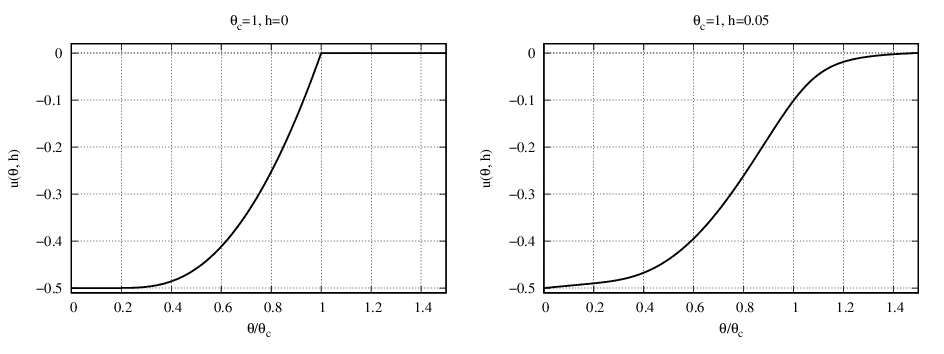}
	\caption{Dependence of the internal energy on $\theta/\theta_c$. Left panel: $h=0$. Right panel $h>0$.} 
	\label{fig:18.7}
\end{figure}

To find the internal energy $U$ we shall use the familiar thermodynamic identity\cite{huang1987}
\[
U=-{{\theta }^{2}}\frac{\partial }{\partial \theta }\left( \frac{F}{\theta } \right)=N{{\theta }^{2}}\frac{\partial }{\partial \theta }{{\left( -\frac{f}{\theta } \right)}_{B}}.
\] 
For the internal energy per one spin this formula gives
\[u=-{{\theta }^{2}}\frac{\partial }{\partial \theta }\left[ -K{{y}^{2}}/2+\ln \cosh \left( Ky+h \right) \right].\] 
As a result of simple calculations we obtain 
\begin{equation}
u\left(\theta,B\right)=-\frac{J}{2}{{y^{2}\left(\theta,B\right)}}-\mu By .
\label{eq:u}
\end{equation}
The first term of this equation is the interaction energy per one spin, whereas the second term is the energy of a spin in the magnetic field. When $\theta = 0$, $y=1$, hence $u=-J/2 - B\mu$. When $B=0$, in the paramagnetic phase $y=0$ and $u=0$. When $B=0$ in the ferromagnetic phase in vicinity of $\theta_c$ one has $K\sim \left(1-\left|t\right|\right)^{-1}$ and
\begin{equation}
y\approx \pm \sqrt{3\left| t \right|/\left( 1-\left| t \right| \right)} ,  
\label{eq:y-vs-t}
\end{equation}
thus, $u\sim \left|t\right|^1$. As we can see from Fig. \ref{fig:18.7}, in the absence of the magnetic field the behaviour of internal energy is singular at $K=1$. The curve representing the function $u\left(\theta,h=0\right)$ consist of two branches. At $\theta=\theta_c$ their derivatives are different. When the magnetic field is turned in this singularity is washed out (Fig. \ref{fig:18.7} right panel). 

\begin{figure}[ht]
	\centering
		\includegraphics[width=0.80\textwidth, angle=0]{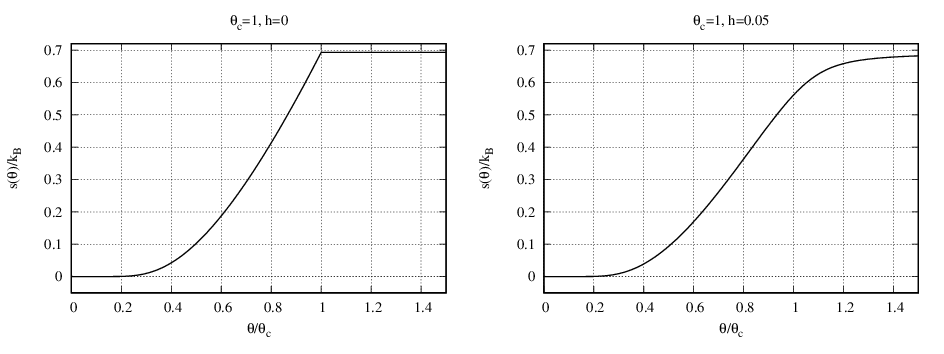}
	\caption{Dependence of entropy on $\theta/\theta_c$. Left panel: $h=0$. Right panel: $h>0$.} 
	\label{fig:18.8}
\end{figure}

Consider entropy 
\begin{equation}
S=-N{{k}_{B}}{{\left( \frac{\partial f\left( \theta ,B \right)}{\partial \theta } \right)}_{B}}.
\label{eq:entropy}
\end{equation}
Calculating the derivative we obtain for entropy per particle the familiar thermodynamic identity \cite{huang1987}
\begin{equation}
s\left(\theta,B\right)=\frac{k_{B}}{\theta}\left[-f\left(\theta,B\right)+u\left(\theta,B\right)\right].
\label{eq:ident-for-entropy}
\end{equation}
For low temperatures $\left(\theta \to 0\right)$
\[
f\left( {\theta ,B} \right) \sim \frac{J}{2}.
\]
Therefore, even when the magnetic is turned in, at $\theta=0$ entropy vanishes. The spins are completely ordered and entropy acquires the lowest value. Since in the absence of the magnetic field the internal energy is singular at $K_c$, entropy is also singular at the critical point. When $B=0$ in the paramagnetic phase ($\theta\geq\theta_c$) $y=0$, and the internal energy vanishes. From the definition \eref{eq:18.12} it follows that $-k_{B}f\left(\theta\geq \theta_{c},B=0 \right)/\theta =k_{B}\ln 2$, and $s=k_{B}\ln2\simeq 0.7\times {{k}_{B}}$. Spins in the paramagnetic phase are completely disordered and entropy reaches its greatest value. The dependence of entropy on temperature is shown in Fig. \ref{fig:18.8}.

If $B=0$ in the paramagnetic phase $y=0$, hence $u\left( \theta >{{\theta }_{c}},h=0 \right)=0$. In the ferromagnetic phase and in vicinity of $\theta_c$ according Eq. \eref{eq:y-vs-abs-t} $u\sim \left|t\right|^1$. For $\theta =\theta_{c}$ the internal energy vanishes $u=0$. This means that the internal energy is the continuous function of temperature $u(\theta_{c}^{-})=u(\theta_{c}^{+})=0$, with $\theta_{c}^{\pm}=\theta_{c}\pm\epsilon$, and $0<\epsilon \ll 1$. 

Entropy $s$ also is a continuous function of temperature. To show this property in the case of ferromagnetic phase and vicinity of $\theta_c$ we use Eqs. \eref{eq:18.15} and \eref{eq:y-vs-abs-t} and Tylor's series for $\ln\left[\tanh\left(Ky\right)\right]$. We get 
\[
s\approx k_{B}\left({\ln 2}-3\left|t\right|/2\right)\, .
\] 
We see that, as one may expect, that entropy of the ferromagnetic phase is smaller than entropy of paramagnetic phase and for $\left|t\right|=0$ attains its maximal value. 

The behaviour of heat capacity in vicinity of the critical temperature is more complex. To calculate heat capacity per one particle one can use one of two thermodynamic relations, namely \cite{huang1987}
\begin{equation}
c\left( \theta ,B \right)={{k}_{B}}{{\left( \frac{\partial u\left( \theta ,B \right)}{\partial \theta } \right)}_{B}},
\label{eq:heat-capacity-u}
\end{equation}
or 
\begin{equation}
c\left( \theta ,B \right)=\theta {{\left( \frac{\partial s\left( \theta ,B \right)}{\partial \theta } \right)}_{B}}. 
\label{eq:heat-capacity-s}
\end{equation} 
It is an easy task to show that these identities yield the same result. We shall consider Eq. \eref{eq:heat-capacity-u}. From Eq \eref{eq:u} it follows that heat capacity $c\left( \theta ,B \right)$ depends on derivative ${{\left( \partial y/\partial \theta  \right)}_{B}}$
\[
c=-{{k}_{B}}\theta \left( Ky+h \right){{\left( \frac{\partial y}{\partial \theta } \right)}_{B}} .
\]
Differentiating both sides of Eq. \eref{eq:18.13} with respect to $y$, solving the obtained equation for ${{\left( \partial y\left( \theta ,B \right)/\partial \theta  \right)}_{B}}$ and applying the identity \eref{eq:identity-1}, we obtain an analytic expression for this derivative 
\[
{{\left( \frac{\partial y}{\partial \theta } \right)}_{B}}=-\frac{1}{\theta }\frac{\left( Ky+h \right)}{{{\cosh }^{2}}\left( Ky+h \right)-K}. 
\]
With the help of the above relation we obtain the final form of expression for heat capacity per one particle
\begin{equation}
c={{k}_{B}}\frac{\left( 1-{{y}^{2}} \right){{\left( Ky+h \right)}^{2}}}{\left( 1-K \right)+K{{y}^{2}}}. 
\label{eq:c-final}
\end{equation}

When $B=0$ in the paramagnetic phase ($\theta\geq\theta_c$) magnetization $m$ vanishes, i.e. $y=0$. Hence, $c=0$. Since $\mathop {\lim} \limits_{\theta \to 0} \,y=1$ in this limit $c$ also vanishes. 

\begin{figure}[ht]
	\centering
		\includegraphics[width=0.80\textwidth, angle=0]{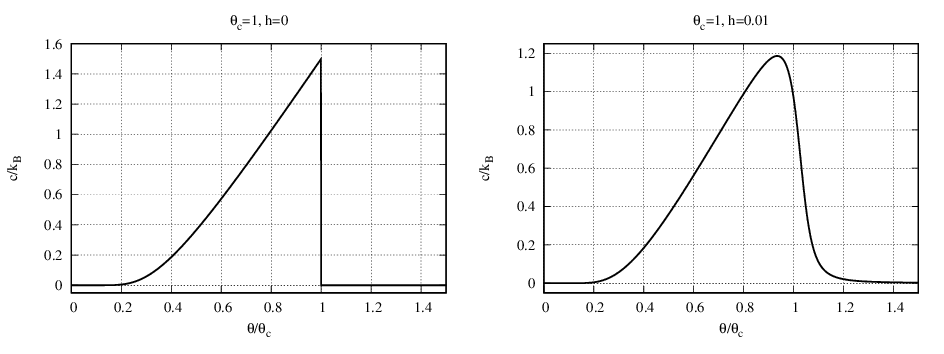}
	\caption{Dependence of heat capacity per one particle on $\theta/\theta_c$. Left panel: $h=0$. Right panel: $h>0$. } 
	\label{fig:18.9}
\end{figure}

Consider heat capacity in ferromagnetic phase in the vicinity of the critical temperature. Using expressions \eref{eq:theta-K-vs-t} and \eref{eq:y-vs-t} we obtain 
\begin{equation}
\frac{c}{{{k}_{B}}}\approx \frac{3}{2}\left(1-\left|t\right|\right). 
\label{eq:c-theta-c}
\end{equation}
Heat capacity is discontinuous at $\theta_c$
\begin{equation}
c\left( \theta _{c}^{-},B=0 \right)-c\left( \theta _{c}^{+},B=0 \right)=3{{k}_{B}}/2.
\label{eq:c-discont}
\end{equation}
The zero field heat capacity is singular at the critical temperature, whereas in the magnetic field it exhibits a peak at the transition point (cf. Fig. \ref{fig:18.9}). 
\section{The magnetic limiting field}
\label{sc:8}
Until now we described the phase transition using the approximate expression for ${\rm{artanh}}~y$. Now we shall give up this approximation, which means that we shall rely on numerical calculations. Such approach will shed light on the approximation used in Sect. \ref{sc:6}. To the best of our knowledge, ours is the first systematic study of limiting magnetic field of the Curie-Weiss model. 

Let us rewrite Eq. \eref{eq:31} in the form 
\begin{equation}
{\rm arctanh}~y-Ky=h. 
\label{eq:31a}
\end{equation}

The function $\Lambda\left(y\right)=\left({\rm{artanh}}~y-Ky\right)$ has two extreme points at $y=\pm \left|t\right|^{1/2}$. For these values of $y$ the limiting value of the parameter $h$ is equal to 
\[{{h}_{t}}=\pm \left[{\rm{arctanh}}\left( \sqrt{\left| t \right|} \right)\mp\frac{\sqrt{\left| t \right|}}{1-\left| t \right|} \right]\,\,,\,\,\left( 0\le T \le {{T}_{c}} \right).\]

\begin{figure}[ht]
	\centering
				\includegraphics[width=0.80\textwidth, angle=0]{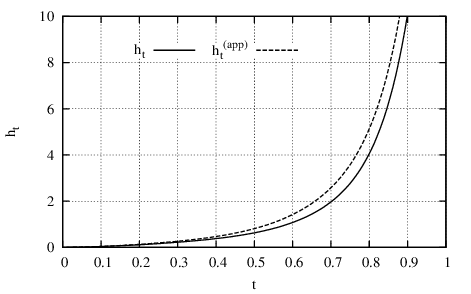}
	\caption{Plot of dependence of critical value of the magnetic field $h_t$ on the parameter $t$. The full line -- result numerical calculation, the dashed line represents the dependence resulting from the cubic equation \eref{eq:18.29}.} 
	\label{fig:18.11}
\end{figure}

For $\left| t \right|\ll 1$ with the accuracy to terms proportional to $\left| t\right|^{3/2}$ the limiting value of $h$ is equal to $h^{(app)}_t$ \eref{eq:h_t} (cf. Fig. \ref{fig:18.11}). 

\begin{figure}[ht]
	\centering
		\includegraphics[width=0.60\textwidth, angle=0]{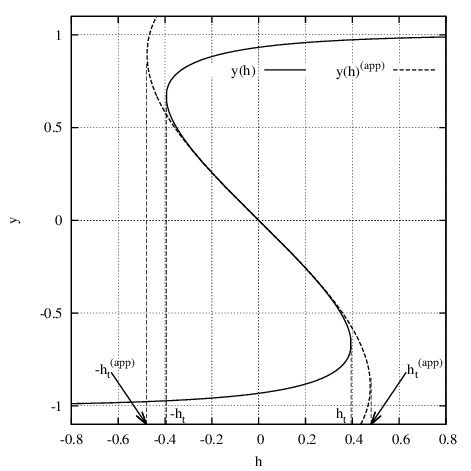}
	\caption{Dependence of $y_{t}\left(h\right)$ on the magnetic field $h$. Dashed line represents the dependence of roots of Eq. \eref{eq:18.29} on the magnetic field. The full line is plot of solution of Eq. \eref{eq:31a}.} 
	\label{fig:18.10}
\end{figure}

For a given value of $\left|t\right|$ we numerically solve Eq. (\ref{eq:31a}) for various values of $h$. We plot $y=y_{t}\left(h\right)$ in Fig. \ref{fig:18.10}. The obtained plot resembles the plot obtained for roots of Eq. \eref{eq:18.29}. However, the critical values of the parameter $h_{t}$ are smaller than $h_{t}^{\left(app\right)}$ given by Eq. (\ref{eq:h_t}). Besides, unlike for the approximate theory, values of $y_{num}$ are confined in the interval [-1,1].

In Fig. \ref{fig:18.11} we compare values of $h_{t}^{(app)}$ and $h_{t}$. It is seen that for small values of $\left|t\right|$ both plots differ a little. With growing $\left|t\right|$ difference is more pronounced. 

\section{Appendix}
\label{sc:appendix}
We shall show that there exist solutions $y\left(\theta,B\right)$ of Eq. \eref{eq:31} for which the second derivative of function $f_{K,h}\left(y\right)$ (Eq. \eref{eq:f_Kh}) with respect to $y$ calculated at $y=y\left(\theta,B\right)$
\begin{equation}
{{\left. {{\left( \frac{{{\partial }^{2}}f_{\theta,h}\left(y\right)}{\partial {{y}^{2}}} \right)}_{\theta ,h}} \right|}_{y=y\left( \theta ,B \right)}}=-{\theta }K\left\{ K\left[ 1-{{y}^{2}}\left( \theta ,B \right) \right]-1 \right\}
\label{eq:2-d derivative f_kh}
\end{equation}
is positive. 

Consider temperatures higher than $\theta_c$.
\begin{enumerate}
\item For $B=0$ in the paramagnetic phase $K\leq 1$ and magnetization vanishes. In this case $y=0$, therefore, the second derivative \eref{eq:2-d derivative f_kh} $\left[-{\theta }K\left( K-1 \right)\right]>0$ is positive.
\item If $B>0$ magnetization is positive, hence $1>y>0$. The double inequality $0<K\left[ 1-{{y}^{2}}\left( \theta ,B \right) \right]<1$ holds. Therefore, $K\left[ 1-{{y}^{2}}\left( \theta ,B \right) \right]-1<0$. Hence, the the second derivative \eref{eq:2-d derivative f_kh} is positive. The same arguments are valid for the negative value of induction ($B<0$ and $y<0$).
\end{enumerate}

In the case of $\theta<\theta_{c}$ the parameter $K$ is greater than unity. As we know (cf. Sect. \ref{sc:4}), if $B=0$, Eq. \eref{eq:18.13} has three solutions, namely $y=0$ and $y=\pm y\left(\theta\right)\equiv \pm y_0$. 
\begin{enumerate}
\item For $y=0$ the derivative \eref{eq:2-d derivative f_kh} is negative
\begin{equation}
{{\left. {{\left( \frac{{{\partial }^{2}}{{f}_{\theta ,h}}\left( y \right)}{\partial {{y}^{2}}} \right)}_{\theta ,h}} \right|}_{y=0}}={\theta }K\left(1- K\right)<0 .
\label{eq:58}
\end{equation}
This means that in the ferromagnetic phase the solution $y=0$ corresponds to a~maximum of free energy. 
\item In the case of two remaining solutions the second derivative reads
\begin{equation}
{{\left. {{\left( \frac{{{\partial }^{2}}{{f}_{K,h}}\left( y \right)}{\partial {{y}^{2}}} \right)}_{\theta ,h}} \right|}_{y=\pm{{y}_{0}}}}=-{\theta }K\left[ K\left( 1-y_{0}^{2} \right)-1 \right]. 
\label{eq:2-nd deriv-feerro-B=0}
\end{equation}
We shall express the parameter $K$ by $y_0$. Using Eq. \eref{eq:31} we can write 
\[
K=\frac{{\rm arctanh}{{y}_{0}}}{{{y}_{0}}}. 
\]
With the help of this relation we find 
\begin{equation}
{{\left. {{\left( \frac{{{\partial }^{2}}{{f}_{K,h}}\left( y \right)}{\partial {{y}^{2}}} \right)}_{\theta ,h}} \right|}_{y=\pm{{y}_{0}}}}=-{\theta }K\left[ \frac{{\rm arctanh}{{y}_{0}}}{{{y}_{0}}}\left( 1-y_{0}^{2} \right)-1 \right]\,.
\label{eq:2-nd-deriv-ferro}
\end{equation}
For small $y_0$ we can use the approximate expression ${\rm arctanh}{{y}_{0}}\approx {{y}_{0}}+y_{0}^{3}/3$. This yields the inequality 
\[
\frac{{\left( 1-y_{0}^{2} \right){\rm arctanh}{y}_{0}}}{y_0}\approx \left( 1+y_{0}^{2}/3 \right)\left( 1-y_{0}^{2} \right)<\left( 1-y_{0}^{4} \right)<1, 
\] 
and the second derivative of free energy is positive. In the ferromagnetic phase, in vicinity of critical temperature, for solutions $\pm y_0$ free energy reaches a minimum. 

The function 
\[
\varphi \left( {{y}_{0}} \right)=\frac{{\rm arctanh}{{y}_{0}}}{{{y}_{0}}}\left( 1-y_{0}^{2} \right) 
\]
is monotonically decreasing in the interval $\left( 0,\left. 1 \right] \right.$ and $0\le \varphi \left( {{y}_{0}} \right)<1$.

\begin{figure}[htb]
	\centering
		\includegraphics[width=0.80\textwidth, angle=0]{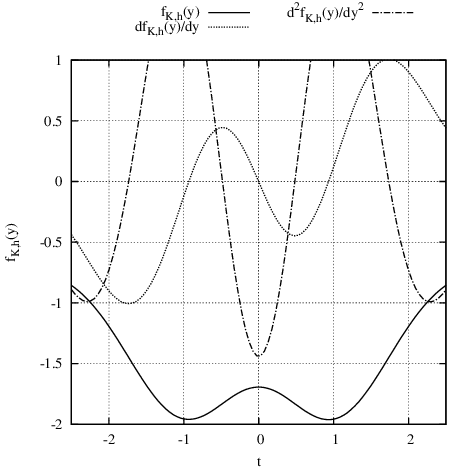}
	\caption{Plot of $f_{K,h}\left(y\right)$ (solid line), $df_{K,h}\left(y\right)/dy$ (dotted line) and $d^{2}f_{K,h}\left(y\right)/dy^2$ (chain line). To zeros of $df_{K,h}\left(y\right)/dy$ there correspond extremums of $f_{K,h}\left(y\right)$. Two of them are minimums because $\frac{{{d}^{2}}{{f}_{K,h}}\left( y \right)}{d{{y}^{2}}}>0$. }
	\label{fig:18.12}
\end{figure}
Using the logarithmic representation \cite{abramowitz1964} ${\rm arctanh}y_{0}=\frac{1}{2}\ln\left(\frac{y_{0}+1}{y_{0}-1}\right)$ and the limiting value \cite{abramowitz1964} ${\lim_{x\to 0} }\,x\ln x=0$  we can show that 
\[
{\lim_{{y}_{0}\to {{1}^{-}}} }\,\frac{{\rm arctanh}{{y}_{0}}}{{{y}_{0}}}\left( 1-y_{0}^{2} \right)=0, 
\]
hence, for low temperatures the right hand side of Eq. \eref{eq:2-nd-deriv-ferro} is positive. 
The plot of function \eref{eq:2-nd-deriv-ferro} is shown in Fig. \ref{fig:18.12}. We conclude that in the ferromagnetic phase minimums of free energy occur for $y=\pm y_0$. 
\item Assume that $B>0$ ($h>0$) and $\theta < \theta_c$ ($K>1$). In this case, according to Sect. \ref{sc:5}, Eq. (\ref{eq:18.13}) has three solutions $y_{1}^{\left( < \right)}\left( t,h \right)>0,\,\,y_{2}^{\left( <\right)}\left( t,h \right)<0,\,\,y_{3}^{\left( < \right)}\left( t,h \right)<0$. Proceeding as before we express $K$ by $y_{1}^{\left( < \right)}$ and $h$  
\begin{equation}
K=\frac{{\rm arctanh}y_{1}^{\left( < \right)}-h}{y_{1}^{\left( <\right)}}. 
\label{eq:K_vs_y_h}
\end{equation}
Now, Eq. \eref{eq:2-d derivative f_kh} takes the form 
\begin{equation}
{{\left. {{\left( \frac{{{\partial }^{2}}{{f}_{K,h}}\left( y \right)}{\partial {{y}^{2}}} \right)}_{\theta ,B}} \right|}_{y=y_{1}^{\left(<\right)}}}=-{\theta }K\left\{ \frac{ {\rm arctanh}y_{1}^{\left( <\right)}-h }{y_{1}^{\left( < \right)}}\left[ 1-{{\left( y_{1}^{\left( < \right)} \right)}^{2}} \right]-1 \right\} .
\label{eq:ineq_y_0}
\end{equation}
Since $0<y_{1}^{\left(<\right)}\le 1$, the inequality $\left[ 1-{{\left( y_{1}^{\left(<\right)} \right)}^{2}} \right]<1$ holds. Thus, the additional term of Eq. \eref{eq:ineq_y_0} is non-positive 
\[
-\frac{h}{y_{1}^{\left( < \right)}}\left[ 1-{{\left( y_{1}^{\left( < \right)} \right)}^{2}} \right]\le 0,
\]
and, even in this case, the second derivative of free energy is positive. 
\end{enumerate}


\end{document}